\documentclass[
reprint,
 amsmath,
 amssymb,
 aps,
 pra,
]{revtex4-2}

\usepackage[utf8]{inputenc}
\usepackage{xcolor}
\usepackage[colorlinks=true,citecolor=blue]{hyperref}
\usepackage{graphicx}
\usepackage{subfigure}
\usepackage{bookmark}
\usepackage[normalem]{ulem}
\usepackage{comment}
\usepackage{dsfont} 
\usepackage[ruled,linesnumbered]{algorithm2e}
\usepackage{physics}
\usepackage{xfrac}
\usepackage{here}
\usepackage{float}
\usepackage{hyperref}

\renewcommand{\d}{\mathrm{d}}
\newcommand{\comma}{~,}
\newcommand{\fullstop}{~.}
\newcommand{\omtilde}{\tilde{\omega}}
\newcommand{\phitilde}{\tilde{\varphi}}
\renewcommand{\vec}[1]{\mathbf{#1}}
\newcommand{\1}{\mathds{1}}
\newcommand{\nstep}{n_\mathrm{step}}

\begin{document}

\title{Geometric Phase in Quantum Synchronization}

\author{Aaron Daniel$^{1}$}
\author{Christoph Bruder$^{1}$}
\author{Martin Koppenh{\"o}fer$^{2}$}
\affiliation{
$^1$ Department of Physics, University of Basel, Klingelbergstrasse 82, CH-4056 Basel, Switzerland\\
$^2$ Pritzker School of Molecular Engineering, University of Chicago, Chicago, Illinois 60637, USA}
\date{\today}

\begin{abstract}
We consider a quantum limit-cycle oscillator implemented in a spin system whose quantization axis is slowly rotated.
Using a kinematic approach to define geometric phases in nonunitary evolution, we show that the quantum limit-cycle oscillator attains a geometric phase when the rotation is sufficiently slow. 
In the presence of an external signal, the geometric phase as a function of the signal strength and the detuning between the signal and the natural frequency of oscillation shows a structure that is strikingly similar to the Arnold tongue of synchronization. 
Surprisingly, this structure vanishes together with the Arnold tongue when the system is in a parameter regime of synchronization blockade. 
We derive an analytic expression for the geometric phase of this system, valid in the limit of slow rotation of the quantization axis and weak external signal strength, and we provide an intuitive interpretation for this surprising effect. 
\end{abstract}

\maketitle

\section{Introduction}

In a seminal paper, Berry showed that a quantum system initialized in an eigenstate of its parameter-dependent Hamiltonian acquires a so-called geometric phase factor upon adiabatic transport around a closed path in parameter space~\cite{Berry1984}. 
Unlike the familiar dynamical phase, acquired by the system due to its time evolution, the geometric phase (GP) depends solely on the curvature of the parameter space and the path taken through  it, and is thereby a purely geometric quantity. 
Pancharatnam had already discovered a similar phase in classical optics earlier~\cite{Pancharatnam1956} and Hannay generalized the concept to classical mechanics afterwards~\cite{Hannay1985}. 
A prominent example of a GP in classical mechanics is provided by the Foucault pendulum, whose plane of oscillation rotates by an angle that depends only on the latitude of the pendulum if the period of oscillation is much shorter than a day~\cite{Shapere1989}. 
GPs appear in diverse settings including light propagation in an optical fiber~\cite{Tomita1986}, the Aharonov-Bohm 
effect~\cite{Ehrenberg1949,Aharonov1959,AharonovAnandan1987}, and the quantum Hall effect~\cite{Arovas1984}.
GPs have also been proposed~\cite{Zanardi1999,Duan2001} as a way to implement quantum gates that are robust against certain pulse imperfections and parameter 
uncertainties~\cite{Filipp2009,Johansson2012,Berger2013}, and such gates have been experimentally demonstrated in a number of systems~\cite{Jones2000,Hua2013,Toyoda2013,Kleissler2018,Huang2019}.

Building on the work of Pancharatnam, the concept of a quantum GP has been extended to nonadiabatic evolution~\cite{AharonovAnandan1987}, noncyclic evolution~\cite{Samuel1988}, and mixed states~\cite{Uhlmann1986,Sjoqvist2000,Singh2003}
(including periodic fermionic systems in mixed states \cite{Bardyn2018}).
A so-called kinematic approach has been formulated~\cite{MUKUNDA1993205}, which is based on the (time dependent) density matrix along a path in state space and enables the concept of a GP to be generalized to nonunitary evolution~\cite{Tong2004}. 
In classical nonlinear dynamics, Kepler \emph{et al.}~\cite{Kepler1991a,Kepler1991b} extended the concept of a GP to classical dissipative systems with self-sustained oscillations (i.e., limit cycles). 
They showed that, even though these systems are nonconservative, cyclic adiabatic deformations of the limit cycle lead to GP shifts which could potentially be observed in certain chemical reactions.

In this paper, we analyze geometric phases in quantum limit-cycle oscillations using the definition of a GP in nonunitary quantum evolution proposed in Ref.~\onlinecite{Tong2004}.
We implement a numerically stable algorithm to calculate the GP of a quantum system undergoing dissipative time evolution. We use this algorithm to demonstrate that a quantum limit-cycle oscillator implemented in a spin-$1$ system acquires a GP that depends only on the trajectory through parameter space if the direction of its quantization axis changes slowly compared to the timescales of its coherent and dissipative dynamics, similar to the classical case~\cite{Kepler1991a,Kepler1991b}.

We then consider the more general case of a quantum limit-cycle oscillator subject to an external signal and show that the GP has a tongue-like structure very similar to the well-known Arnold tongue in synchronization~\cite{Pikovsky2003}.
Surprisingly, this tongue-like structure of the GP vanishes together with the conventional Arnold tongue of synchronization if the system is in a regime of interference-based quantum synchronization blockade~\cite{Koppenhoefer2019}. 
To understand this surprising effect, we derive an analytical formula for the GP, which is valid if the quantization axis is rotated sufficiently slowly. 

This paper is structured as follows:  In Sec.~\ref{sec:gp_in_open_qs}, we summarize the kinematic approach to the GP in an open quantum system before we introduce the numerical algorithm to compute the GP in a generic open quantum system in Sec.~\ref{sec:NumericalAlgorithm}. 
In Sec.~\ref{sec:gp_quantum_lc}, we focus on the specific example of a  van der Pol (vdP) oscillator subject to a weak external signal.  We demonstrate the surprising similarities between the GP and the Arnold tongue of synchronization in this system. 
To gain better insight into this phenomenon, we derive an analytical expression for the GP in an arbitrary quantum limit-cycle oscillator with nondegenerate populations. 
Finally, we conclude in Sec.~\ref{sec:conclusion}.

\section{Geometric phase in open quantum systems}
\label{sec:gp_in_open_qs}

The GP of a quantum system can be defined as the difference between the global phase acquired during the time evolution and the local phase changes accrued along the way~\cite{Sjoqvist2015}. 
This subtraction is equivalent to enforcing a parallel-transport condition~\cite{Sjoqvist2015}.
To apply this definition to a quantum system in a mixed state undergoing nonunitary evolution, one has to consider the GP of a purification of the system, 
which is measurable in principle~\cite{Tong2004} by an interferometric measurement of the purified state. 
Note, however, that the interferometric measurement requires unitary evolution of an enlarged system comprising the system of interest and ancillary degrees of freedom because the value of the GP depends on the chosen purification~\cite{Rezakhani2006} (see App.~\ref{sec:App:mzi_measurement} for more details). 
Given an open quantum system undergoing evolution along a path 
$\mathcal{P}$ in the space of density matrices,
\begin{align}
	\mathcal{P} : t \in [0,\tau] \mapsto \hat{\rho}(t) = \sum_{k=1}^N p_k(t) \ket{\phi_k(t)} \bra{\phi_k(t)} \comma \label{eq:pathtong}
\end{align}
where $\hat{\rho}(t)$ is the time-dependent density matrix of the system, $p_k \geq 0$ are its populations (which we assume to be nondegenerate functions for $t \in [0,\tau]$), $\ket{\phi_k(t)}$ are the corresponding eigenvectors, and $N$ is the Hilbert-space dimension, Tong \emph{et al.}~\cite{Tong2004} proposed
the following definition of the GP $\gamma[\mathcal{P}]$:
\begin{align}
	\gamma[\mathcal{P}] = \arg \Bigg[ \begin{aligned}[t]
         &\sum_{k=1}^N \sqrt{p_k(0) p_k(\tau)} \bra{\phi_k(0)} \ket{\phi_k(\tau)} \\
         &\times \exp(- \int_0^\tau \bra{\phi_k(t)} \ket{\dot{\phi}_k(t)} \d t)  \Bigg] \fullstop
    \end{aligned}
    \label{eq:GPtong}
\end{align}
Intuitively, Eq.~\eqref{eq:GPtong} is the sum over the Pancharatnam phases of each eigenstate $\ket{\phi_k(t)}$ of the density matrix $\hat{\rho}(t)$, weighted by the corresponding population $p_k(t)$ at the start and the end of the path $\mathcal{P}$, where the exponential factors subtract the local phase changes accrued along $\mathcal{P}$.

\section{Algorithm to calculate the geometric phase numerically}
\label{sec:NumericalAlgorithm}

Solutions to dissipative quantum systems in closed analytical form are rare: beyond the simplest models, it is impossible to solve the differential equations arising from a quantum master equation (QME) analytically. 
We therefore use a numerical approach to evaluate Eq.~\eqref{eq:GPtong}
and calculate the time evolution of a given quantum system by solving its QME numerically exactly using the \textsc{QuantumOptics} package~\cite{kramer2018quantumoptics} in \textsc{Julia}~\cite{Bezanson2017}.
This provides us with the density matrix $\hat{\rho}_j = \hat{\rho}(t_j)$ of the system at equidistant discrete time steps $t_j = j \Delta t$, $j \in \{0, \dots, \nstep \}$, where $\Delta t = \tau/\nstep$. Using this set of density matrices $\{\hat{\rho}_j\}$ we then evaluate Eq.~\eqref{eq:GPtong} using Algorithm ~\ref{alg:gp_algorithm}.

\begin{algorithm}[t]
\SetAlgoLined
	\SetKwInOut{Input}{Input}\SetKwInOut{Output}{Output}
	\Input{$\{\hat{\rho}_j \mid j = 0, \dots, \nstep \}$ where $\hat{\rho}_j = \hat{\rho}(t_j)$ and $t_j = j \Delta t$.}
	\Output{Geometric phase $\gamma[\mathcal{P}]$ defined in Eq.~\eqref{eq:GPtong}.}
	Compute the eigenstates $\ket{\phi_k(t_j)}$ and populations (statistical weights) $p_{k}(t_j)$ for each $\hat{\rho}_j$ and apply a phase convention\label{alg:step1}\;
	Differentiate the eigenstates to obtain $\ket{\dot{\phi}_k(t_j)}$\label{alg:step2}\; 
	Integrate $\bra{\phi_k(t_j)} \ket{\dot{\phi}_k(t_j)}$ over time\label{alg:step3}\; 
	Evaluate Eq. \eqref{eq:GPtong} using the results of steps \ref{alg:step1} and \ref{alg:step3}.
	\caption{
		Numerical calculation of the geometric phase using Eq.~\eqref{eq:GPtong}.
	}
	\label{alg:gp_algorithm}
\end{algorithm}

For each time step $t_j$, we diagonalize $\hat{\rho}_j$ numerically to find the populations $p_k(t_j)$ and the associated eigenvectors $\ket{\phi_k(t_j)}$ defined in Eq.~\eqref{eq:pathtong}. 
Note that for the specific problem considered later, the populations $p_k(t)$ are constant and distinct, such that we can order them ascendingly and there is no ambiguity in the labeling of the eigenvectors in different time steps.
Since the eigenvectors are only defined up to a global phase factor, we use the convention that the $k$th entry of the $k$th eigenvector is real and positive. 
This is equivalent to choosing a particular gauge and does not affect the GP since Eq.~\eqref{eq:GPtong} is gauge-invariant.

In step~\ref{alg:step2} of Alg.~\ref{alg:gp_algorithm}, the exponential phase factors in Eq.~\eqref{eq:GPtong} are calculated by numerically differentiating $\{ \ket{\phi_k(t_j)} \mid j = 1, \dots, \nstep\}$ with respect to time using a symmetric difference quotient that is of fourth order in $\Delta t$ \cite{LI200529}. 
The overlaps $\{\bra{\phi_k(t_j)}\ket{\dot{\phi}_k(t_j)} \mid j = 1, \dots, \nstep\}$ are then numerically integrated over time using an extended Simpson rule in step \ref{alg:step3}, which is also of fourth order in $\Delta t$~\cite{numerical_recipes}.
The specific choice of the gauge of the eigenvectors $\ket{\phi_k(t_j)}$ ensures that the overlaps are smooth functions of time and that the numerical integration is stable.

We benchmarked this algorithm using the exactly solvable case of a qubit subject to dephasing~\cite{Tong2004}, see App.~\ref{sec:App:QubitDephasingBenchmark} for more details.

\section{Geometric phase of a quantum limit-cycle oscillator}
\label{sec:gp_quantum_lc}

\subsection{Quantum van der Pol oscillator}

\begin{figure}
    \centering
    \subfigure[]{
	    \includegraphics[width=0.2\textwidth]{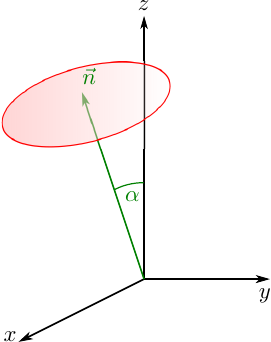}
	    \label{fig:spin1_rot}
    }
    \subfigure[]{
        \includegraphics[width=0.25\textwidth]{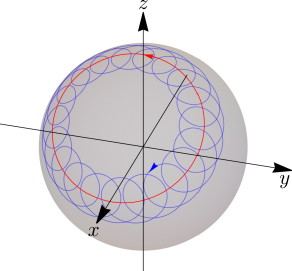}
    		\label{fig:precession_plot}
	}
    \caption{
    (a) We consider a quantum limit-cycle oscillator, implemented in a spin-$1$ system, whose quantization axis is rotated with angular frequency $\omega$ on the surface of a cone with opening angle $\alpha$ and symmetry axis $\vec{n}(\alpha)$, as defined in Eq.~\eqref{eq:Rot}. The quantization axis rotates counter-clockwise about $\vec{n}(\alpha)$ and points along the $z$-direction at time $t=0$. (b) Trajectory of the instantaneous eigenstate $\ket{\phi_{+1}(t)}$ of the density matrix $\hat{\rho}(t)$ on the spin-$1$ Bloch sphere in the laboratory frame, obtained by numerical integration of Eq.~\eqref{eq:QME:LabFrame}. In the limit-cycle state, i.e., without any external signal, the eigenstate follows the rotation of the quantization axis (red line). If an external signal $\hat{H}_\mathrm{sig}(t) \neq 0$ is applied to synchronize the quantum limit-cycle oscillator, $\ket{\phi_{+1}(t)}$ precesses about the instantaneous quantization axis (blue line). The eigenstate $\ket{\phi_{-1}(t)}$ follows a similar trajectory on the opposite part of the Bloch sphere (not shown). For presentation purposes, $\omega$ and the signal strength $T$ have been chosen much larger than in the numerical examples.}
    \label{fig:sketches}
\end{figure}

The numerical algorithm introduced in Sec.~\ref{sec:NumericalAlgorithm} can be applied to any quantum system whose density matrix $\hat{\rho}(t)$ is known as a function of time and has distinct populations $p_k(t)$. 
In the following, we focus on a specific example, namely, a quantum limit-cycle oscillator implemented in a spin-$1$ system whose quantization axis changes slowly. 

A classical limit-cycle oscillator is a nonlinear dynamical system with an internal source of energy that excites the system into self-sustained periodic motion at the so-called natural frequency $\omega_0$~\cite{Pikovsky2003}.
The phase of this oscillation is free, such that the limit-cycle oscillator can adjust its frequency in the presence of a weak periodic drive. 
A similar phenomenon occurs if multiple limit-cycle oscillators are coupled, and is called mutual synchronization.
Several proposals have been put forward to generalize the concept of synchronization to the quantum regime~\cite{Zhirov2006,Goychuk2006,Giorgi2012,Hriscu2013,Ludwig2013,Lee2013,Walter2014,Xu2014,DavisTilley2016,Roulet2018,Koppenhoefer2019} and to quantify it~\cite{Ludwig2013,Lee2013,Mari2013,Lee2014,Ameri2015,Hush2015,Weiss2016,Galve2017,Jaseem2020,EshaqiSani2020}.

Here, we consider a quantum limit-cycle oscillator implemented in a spin-$1$ system, which is convenient since it allows us to work with a finite-dimensional Hilbert space, $N=3$. 
We follow the framework introduced in Ref.~\onlinecite{Koppenhoefer2019}, which defines synchronization based on the phase-space dynamics of the spin system. 
A quantum limit-cycle oscillator can be modeled by a QME of the form ($\hbar = 1$)
\begin{align}
	\frac{\d}{\d t} \hat{\rho} = - i \commutator{\hat{H}_0}{\hat{\rho}} + \sum_{j=1}^M \mathcal{D}[\hat{\Gamma}_j] \hat{\rho} \comma
	\label{eq:QME}
\end{align}
where $\mathcal{D}[\hat{O}] = \hat{O} \hat{\rho} \hat{O}^\dagger - \anticommutator{\hat{O}^\dagger \hat{O}}{\hat{\rho}}/2$ is a Lindblad dissipator. 
The Hamiltonian $\hat{H}_0$ determines the natural frequency of oscillation $\omega_0$ of the limit-cycle oscillator. 
We choose the quantization axis to be the $z$-direction and set
\begin{align}
	\hat{H}_0 = \omega_0 \hat{S}_z \fullstop
	\label{eq:H0}
\end{align}
The spin operators obey the commutation relation $[\hat{S}_j,\hat{S}_k] = i \varepsilon_{jkl} \hat{S}_l$, where $j,k,l \in \{x,y,z\}$, and they are the generators of the rotation group $SO(3)$.
A basis of the Hilbert space is given by the joint eigenstates $\ket{S,m}$ of $\vec{\hat{S}}^2 = \hat{S}_x^2 + \hat{S}_y^2 + \hat{S}_z^2$ and $\hat{S_z}$, where $S=1$ and $m \in \{+1,0,-1\}$.
The set of jump operators $\{ \hat{\Gamma}_1, \dots, \hat{\Gamma}_M\}$ determines how the amplitude of the limit-cycle oscillator is stabilized in phase space and should not introduce any phase preference of the oscillation. For our numerical examples, we consider a spin-$1$ implementation of a quantum vdP oscillator~\cite{Lee2013,Walter2014}, such that $M=2$ and
\begin{align}
	\hat{\Gamma}_1 &= \sqrt{\frac{\gamma_{\mathrm{g}}}{2}} \left( \sqrt{2} \hat{S}_z \hat{S}_+ - \hat{S}_+ \hat{S}_z\right) \comma 
	\label{eq:jump_operators:1}\\
	\hat{\Gamma}_2 &= \sqrt{\frac{\gamma_{\mathrm{d}}}{2}} \hat{S}_-^2 \fullstop
	\label{eq:jump_operators:2}
\end{align}
Here, $\gamma_{\mathrm{g}}$ and $\gamma_{\mathrm{d}}$ denote the gain and damping rates, respectively, and $\hat{S}_\pm = \hat{S}_x \pm i \hat{S}_y$ are the raising and lowering operators.
The specific form of the jump operators~\eqref{eq:jump_operators:1} and~\eqref{eq:jump_operators:2} can be motivated as follows.
In the quantum regime, $\gamma_\mathrm{g} \ll \gamma_\mathrm{d}$, the bosonic quantum vdP oscillator populates only the lowest three Fock states \cite{Lee2013,Walter2014}. 
Thus, the bosonic system can be approximated by a spin-$1$ system whose jump operators have the same matrix representation as the corresponding bosonic jump operators restricted to the lowest three Fock states \cite{Koppenhoefer2019}.

\subsection{Demonstration of a geometric phase in a quantum van der Pol oscillator}
\label{sec:quantumvdP:Undriven}

Kepler \emph{et al.}~\cite{Kepler1991a} demonstrated that a classical limit-cycle oscillator acquires a geometric phase if its limit cycle is adiabatically deformed along a closed path in parameter space. 
To generate a similar effect in the quantum limit-cycle oscillator defined in Eq.~\eqref{eq:QME}, we choose to rotate the direction of the quantization axis along a path $\mathcal{P}_\mathrm{lab}$ in parameter space, i.e.,
\begin{align}
	\hat{H}_0 &\to \hat{R}(\alpha,t) \hat{H}_0 \hat{R}^\dagger(\alpha,t) \comma \\
	\hat{\Gamma}_j &\to \hat{R}(\alpha,t) \hat{\Gamma}_j \hat{R}^\dagger(\alpha,t) \comma
\end{align}
where the rotation operator is  
\begin{align}
	\hat{R}(\alpha,t) = e^{-i \omega t \vec{n}(\alpha) \cdot \vec{\hat{S}}} \fullstop
	\label{eq:Rot}
\end{align}
Here, $\vec{\hat{S}} = (\hat{S}_x,\hat{S}_y,\hat{S}_z)^\top$ is the vector of spin operators, and the unit vector $\vec{n}(\alpha) = ( \sin\alpha, 0 , \cos\alpha)^\top$  defines the symmetry axis of a cone with opening angle $\alpha$. 
The quantization axis rotates on the surface of this cone, as shown in Fig.~\ref{fig:spin1_rot},
which constitutes the path $\mathcal{P}_\mathrm{lab}$ in parameter space. 
The time evolution of the quantum system along the path $\mathcal{P}_\mathrm{lab}$ induces the path $\mathcal{P}$ in the space of density matrices defined in Eq.~\eqref{eq:pathtong}, for which we can calculate the geometric phase $\gamma[\mathcal{P}]$ using Eq.~\eqref{eq:GPtong}. 
As discussed by Aharonov and Anandan \cite{AharonovAnandan1987} in the context of unitary evolution, $\gamma[\mathcal{P}]$ can be viewed equally well as a geometric phase of the path $\mathcal{P}_\mathrm{lab}$ in parameter space in the limit of adiabatic evolution.
\begin{figure}
    \centering
    \includegraphics{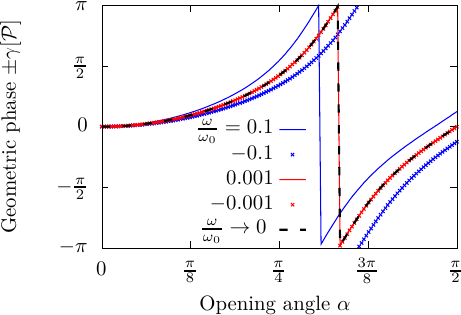}
    \caption{Demonstration of a geometric phase (GP) in a quantum van der Pol limit-cycle oscillator whose quantization axis is slowly rotated on the surface of a cone with opening angle $\alpha$ along the path $\mathcal{P}_\mathrm{lab}$,as shown in Fig.~\ref{fig:sketches}. If the rotation frequency $\omega$ is much smaller than the natural frequency of oscillation $\omega_0$ and the dissipation rates $\gamma_\mathrm{g,d}$, the system acquires a purely geometric phase whose sign depends only on the direction of the path $\mathcal{P}_\mathrm{lab}$ in parameter space (red lines and markers). For larger $\omega$, the evolution is no longer adiabatic (blue lines and markers) and the GP for the respective path $\mathcal{P}$ in the space of density matrices depends also on the velocity at which $\mathcal{P}_\mathrm{lab}$ is traversed. The black dashed line shows an analytical formula for the GP in the limit $\omega/\omega_0 \to 0$, given by Eq.~\eqref{eq:GP:approximateCyclic}. The parameters are  $\omega_0/\gamma_\mathrm{d}= 10$, $\gamma_\mathrm{g}/\gamma_\mathrm{d} = 0.1$, and $\nstep = 10^8$.} \label{fig:no_signal_GP}
\end{figure}
In our scenario, adiabatic evolution corresponds to a rotation frequency $\omega$ of the quantization axis that is much smaller than the relaxation rates and the natural frequency of the system, $\omega \ll \gamma_\mathrm{g},\gamma_\mathrm{d}, \omega_0$~\cite{Lidar2005}.
To demonstrate the existence of a GP $\gamma[\mathcal{P}]$, 
which (in addition to being a geometric quantity of the path $\mathcal{P}$ in the space of density matrices) depends only the geometry of the path  $\mathcal{P}_\mathrm{lab}$ through parameter space,
we simulate a full rotation of the quantization axis for different values of the rotation frequency $\omega$, and compute the phase acquired along this path $\mathcal{P}_\mathrm{lab}$ with Alg.~\ref{alg:gp_algorithm}. 
For slow rotation of the quantization axis, i.e., adiabatic evolution, we expect the resultant phase to depend only on the geometry of the path  traced out by the rotation. Therefore, reversing the rotation direction, $\mathcal{P}_\mathrm{lab} \to - \mathcal{P}_\mathrm{lab}$, should only flip the sign of the phase. 
Figure~\ref{fig:no_signal_GP} demonstrates that 
this is indeed the case if 
the rotation frequency $\omega$ is small enough, 
thus, the limit-cycle oscillator indeed acquires a GP in the regime $\omega \ll \gamma_\mathrm{g},\gamma_\mathrm{d},\omega_0$ 
that 
depends
only on the rotation of the quantization axis in parameter space. 
Note that, in the nonadiabatic case (i.e., for fast rotations of the quantization axis compared to the relaxation rates and natural frequency of the system), the geometric phase along the path $\mathcal{P}$ can no longer be viewed as a geometric quantity of the path $\mathcal{P}_\mathrm{lab}$ in parameter space, such that the geometric phases obtained for $\mathcal{P}_\mathrm{lab}$ and $-\mathcal{P}_\mathrm{lab}$ differ.
These numerical 
results in Fig.~\ref{fig:no_signal_GP}
can be explained by the following heuristic argument.
In the limit of infinitely slow rotation, i.e., $\omega \to 0$, we expect  the system always to remain in the steady state along the current direction of the quantization axis, i.e., $\hat{\rho}(t)$ is well approximated by the steady-state solution of Eq.~\eqref{eq:QME} rotated by $\hat{R}(\alpha,t)$.
Sj\"oqvist \emph{et al.}~\cite{Sjoqvist2000} showed that the GP of a mixed state $\hat{\rho}(t)$ undergoing unitary evolution is the weighted sum of the GPs acquired by each eigenvector $\ket{\phi_k(t)}$, i.e., we find 
\begin{align}
	\gamma[\mathcal{P}] = \arg \left[ p_{+1} e^{+2 \pi i \cos\alpha} + p_0 + p_{-1} e^{-2 \pi i \cos\alpha} \right] \comma 
	\label{eq:GP:CyclicNoSignal}
\end{align}
where the populations are
\begin{align}
	p_{+1} &= \frac{\gamma_\mathrm{g}}{3 \gamma_\mathrm{d} + \gamma_\mathrm{g}} \comma 
	&
	p_{0}  &= \frac{\gamma_\mathrm{d}}{3 \gamma_\mathrm{d} + \gamma_\mathrm{g}} \comma 
	&
	p_{-1} &= \frac{2 \gamma_\mathrm{d}}{3 \gamma_\mathrm{d} + \gamma_\mathrm{g}} \fullstop
	\label{eq:vdP:population}
\end{align}
The phase factors given by the exponential functions measure the solid angle traced out by each eigenvector $\ket{\phi_k(t)}$ and are a generalization of Berry's result to a spin-$1$ system~\cite{Berry1984}.
As shown in Fig.~\ref{fig:no_signal_GP}, this formula for the GP in the limit $\omega \to 0$ matches perfectly with the numerical results.
Note that we will provide a more rigorous derivation of the GP in Sec.~\ref{sec:approx_expr_gp}, where we show that the heuristically motivated result given by Eqs.~\eqref{eq:GP:CyclicNoSignal} and~\eqref{eq:vdP:population} is a limiting case of a more general calculation.

\subsection{Arnold tongue of the geometric phase in the presence of an external signal}
\label{sec:quantumvdP:driven}

So far, we have shown that a single isolated quantum limit-cycle oscillator acquires a geometric phase upon adiabatic rotation of its quantization axis. 
Quantum limit-cycle oscillators are of particular interest because they can be synchronized to an external signal at frequency $\omtilde$. 
In this so-called entrainment phenomenon, the external signal causes the limit-cycle oscillator to deviate from its natural frequency $\omega_0$.
The degree to which the frequency of oscillation is modified depends on the detuning $\Delta = \omtilde - \omega_0$ and the strength of the signal.
In general, the entrainment is strongest on resonance, $\Delta = 0$, and the range of detuning where synchronization occurs grows with increasing signal strength $T$. 
This gives rise to the so-called Arnold tongue of synchronization, a roughly triangular-shaped region in the $\Delta$-$T$ parameter space~\cite{Pikovsky2003}.
To describe the presence of an external signal, we replace $\hat{H}_0 \to \hat{H_0} + \hat{H}_\mathrm{sig}(t)$ in Eq.~\eqref{eq:QME} with the signal Hamiltonian
\begin{align}
	\hat{H}_\mathrm{sig}(t) = T \cos (\omtilde t + \phitilde) \hat{S}_x \comma
	\label{eq:Hsig}
\end{align}
which aims to rotate the state of the limit-cycle oscillator about an axis in the equatorial plane at an angle $\phitilde$ with respect to the positive $x$-axis. 
In the remaining parts of this paper, we focus on the GP of a quantum limit-cycle oscillator subject to the external signal given by Eq.~\eqref{eq:Hsig}, and we discover striking similarities between the Arnold tongue of synchronization and a corresponding plot of the GP as a function of $\Delta$ and $T$.

As a preparation, we first calculate the Arnold tongue of synchronization of the system for a fixed quantization axis, i.e., for $\omega = 0$.
In a frame rotating at the signal frequency $\omtilde$ and using a rotating-wave approximation, Eq.~\eqref{eq:QME} becomes time-independent and, to first order in the small signal strength $T$, its steady state has the form

\begin{align}
	\hat{\rho}_\mathrm{ss} = \begin{pmatrix}
		p_{+1} \\ & p_0 \\ & & p_{-1}
	\end{pmatrix} + T \begin{pmatrix}
		& c_{+1,0} \\
		c_{+1,0}^* & & c_{0,-1} \\
		& c_{0,-1}^* 
	\end{pmatrix} \nonumber\\ 
	+ \mathcal{O}(T^2) \fullstop
	\label{eq:sigmasteadystate}
\end{align}
For the quantum vdP oscillator considered here, the populations $p_m$ are given in Eq.~\eqref{eq:vdP:population}  and the coherences $T c_{m,m'}$ are
\begin{align}
	c_{+1,0} &= \frac{a}{b} \comma 
	\label{eq:vdP:cp0} \\ 
    a &= - i e^{-i \tilde{\varphi}} \Big[\begin{aligned}[t]
    		(4 + 3 \sqrt{2}) \gamma_\mathrm{g} \gamma_\mathrm{d} &- 2 \sqrt{2} i \gamma_\mathrm{d} \Delta   \\ 
    		&- \sqrt{2} \gamma_\mathrm{g}(3 \gamma_\mathrm{g} - 2 i \Delta) \Big] \comma 
    \end{aligned} \nonumber \\
    b &= 4 (3 \gamma_\mathrm{d} + \gamma_\mathrm{g})(\gamma_\mathrm{d} + \gamma_\mathrm{g} - i \Delta)(3 \gamma_\mathrm{g} - 2 i \Delta) \nonumber \comma \\
 	c_{0,-1} &= \frac{-i e^{-i \tilde{\varphi}} \gamma_\mathrm{d}}{\sqrt{2} (3 \gamma_\mathrm{d} + \gamma_\mathrm{g})(3 \gamma_\mathrm{g} - 2 i \Delta)} \fullstop 
 	\label{eq:vdP:c0m}
\end{align}
Following Refs.~\onlinecite{Roulet2018} and~\onlinecite{Koppenhoefer2019}, we define a phase-space quasiprobability distribution of the limit-cycle oscillator by calculating its Husimi-Q function $Q(\theta,\phi \vert \hat{\rho}) = \bra{\theta,\phi} \hat{\rho} \ket{\theta,\phi}$, where $\ket{\theta,\phi} = e^{-i \phi \hat{S}_z} e^{- i \theta \hat{S}_y} \ket{S=1,m=+1}$ are coherent spin states~\cite{Radcliffe1971}. 
Since we are working in a frame rotating at $\omtilde$, the variable $\phi$ determines the relative phase between the limit-cycle oscillator and the applied signal. 
From the Q function, we obtain the shifted phase distribution of $\hat{\rho}$,
\begin{align}
    S(\phi|\hat{\rho}) = \int_0^\pi d\theta \sin(\theta) Q(\theta,\phi|\hat{\rho}) - \frac{1}{2\pi} \comma
\end{align}
which is zero if the relative phase is uniformly distributed (no synchronization) and nonzero if there is a preferred relative phase.
A single-number measure of synchronization can be obtained by considering the maximum of $S(\phi \vert \hat{\rho})$, 
\begin{align}
	\mathcal{S}(\hat{\rho}) 
	= \max_{\phi \in [0,2\pi)} S(\phi|\hat{\rho}) 
	= \frac{3}{8 \sqrt{2}} T \abs{c_{1,0} + c_{0,-1}} \fullstop
	\label{eq:SyncMeasure}
\end{align}
$\mathcal{S}$ is positive (zero) if there is (no) synchronization and the particular value of $\phi_\mathrm{max} = - \arg(c_{1,0} + c_{0,-1})$ maximizing $S(\phi \vert \hat{\rho})$ determines the relative phase lag between the limit-cycle oscillator and the signal.

\begin{figure}
    \centering
    \includegraphics[width=0.475\textwidth]{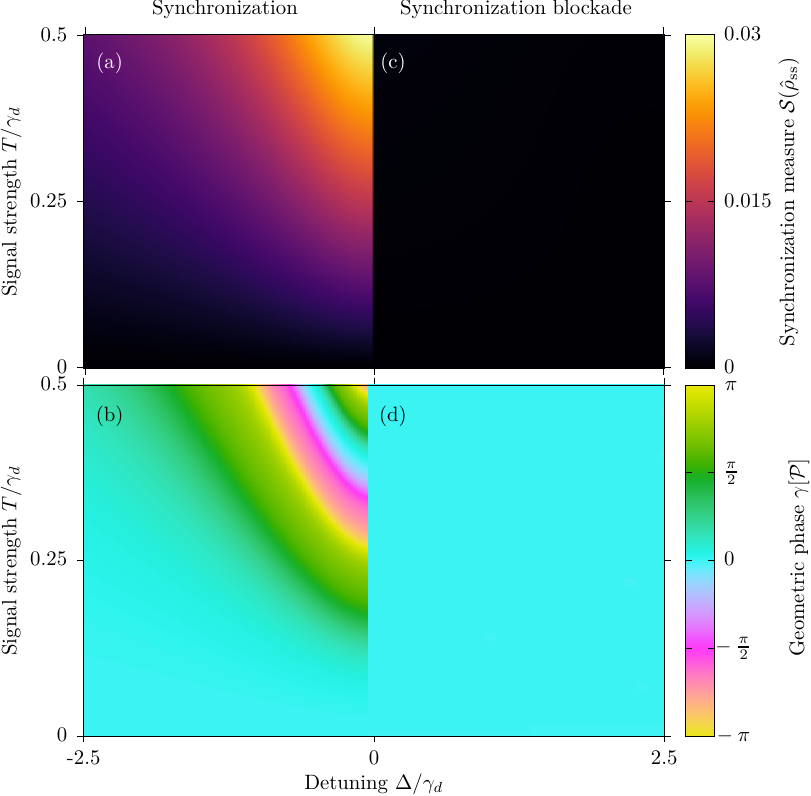}
    \caption{(a) Arnold tongue of a quantum van der Pol oscillator subject to a semiclassical signal given by Eq.~\eqref{eq:Hsig}.
    The limit-cycle oscillator is synchronized to the external signal in the bright region where the synchronization measure $\mathcal{S}(\hat{\rho}_\mathrm{ss})$, defined in Eq.~\eqref{eq:SyncMeasure}, is nonzero. The plot is symmetric about the $\Delta=0$ axis and we only show the left half. The dissipation rates have the ratio $\gamma_\mathrm{g}/\gamma_\mathrm{d}=0.5$.
    (c)The right side of the same plot in a parameter regime where an interference-based quantum synchronization blockade occurs $\qty(\gamma_\mathrm{g}/\gamma_\mathrm{d} \approx 2.84)$. (b) Plot of the GP for the same parameter values as in (a) and $\nstep =10^4$, which shows a strikingly similar Arnold-tongue-like structure. Again, the plot is symmetric about the $\Delta=0$ axis. (d) For the interference-based quantum synchronization blockade parameters of (c) and $\nstep=10^6$, this structure in the GP disappears. For all four cases, the remaining parameters are $\omega_0/\gamma_\mathrm{d}=1,~\tau \omega_0 = 200, ~\tilde{\varphi} = 0,~\alpha = \pi/4,~\omega/\gamma_\mathrm{d} = 0.05$, and $ \omtilde = \omega_0 + \Delta.$ Note that the color scale in (b) and (d) is periodic since the GP is a $2\pi$-periodic quantity (unlike the synchronization measure $\mathcal{S}(\hat{\rho}_\mathrm{ss})$).}
    \label{fig:squareplot}
\end{figure}

The Arnold tongue of quantum synchronization can now be visualized by plotting $\mathcal{S}(\hat{\rho}_\mathrm{ss})$ as a function of $\Delta$ and $T$, which is shown in Fig.~\ref{fig:squareplot}(a). 
In contrast to the classical case, quantum noise smears out the synchronization transition and leads to a smooth crossover from no synchronization at large detunings and weak signal strength (dark colors) to a roughly triangular-shaped region of synchronization for sufficiently large signal strength around resonance (bright colors).

We now analyze the GP of a limit-cycle oscillator with an applied signal whose quantization axis is slowly rotated, $\omega \neq 0$, as shown in Fig.~\ref{fig:spin1_rot}. 
Numerically integrating the QME
\begin{align}
	\frac{\d}{\d t} \hat{\rho} 
	&= - i \commutator{\hat{R}(\alpha,t) \left( \hat{H}_0 + \hat{H}_\mathrm{sig}(t) \right) \hat{R}^\dagger(\alpha,t)}{\hat{\rho}} \nonumber \\
	&\phantom{=}\ + \sum_{j=1}^M \mathcal{D}\left[\hat{R}(\alpha,t) \hat{\Gamma}_j \hat{R}^\dagger(\alpha,t) \right] \hat{\rho} 
	\label{eq:QME:LabFrame}
\end{align}
and using the algorithm described in Sec.~\ref{sec:NumericalAlgorithm} to calculate the GP $\gamma[\mathcal{P}]$, we find the GP shown in Fig.~\ref{fig:squareplot}(b). 
Comparing Figs.~\ref{fig:squareplot}(a) and (b), we find a striking similarity between the GP and the Arnold tongue of synchronization:
Both quantities take a constant value at large detuning and small signal strength, and vary strongly in a triangular region around resonance whose width grows with increasing signal strength.

One may argue that this coincidence is not surprising since both $\mathcal{S}(\hat{\rho})$ and $\gamma[\mathcal{P}]$ depend on the density matrix of the system. 
For $T \to 0$ or $\abs{\Delta} \to \infty$, the external signal cannot significantly affect the limit-cycle oscillator and its density matrix is essentially independent of $T$ and $\Delta$ and equivalent to that of an unperturbed vdP oscillator. 
Close to resonance and for large enough $T$, however, the signal will significantly affect the oscillation dynamics and changes in both the synchronization measure and the GP are to be expected.

However, the similarities between the GP and the synchronization measure do not end here.
In Fig.~\ref{fig:squareplot}(c), we plot $\mathcal{S}(\hat{\rho}_\mathrm{ss})$ for parameters in a so-called interference-based quantum-synchronization-blockade regime~\cite{Koppenhoefer2019}.
In this regime, the gain and damping rates, $\gamma_\mathrm{g}$ and $\gamma_\mathrm{d}$, respectively, are chosen such that the coherences entering the definition of $\mathcal{S}(\hat{\rho})$ in Eq.~\eqref{eq:SyncMeasure} have the same magnitude but opposite signs, $c_{1,0} = - c_{0,-1}$. On resonance, this relation takes the following form 
\begin{align}
    \frac{\gamma_\mathrm{g}}{\gamma_\mathrm{d}} = 4 + 5 \sqrt{2} + \sqrt{48 - \frac{(4+5\sqrt{2})^2}{6\sqrt{2}}} \approx 2.84 \fullstop
\end{align}
In this regime, each coherence is nonzero, i.e., the signal does modify the dynamics of the limit-cycle oscillator appreciably, but an interference effect prevents phase localization such that $\mathcal{S}(\hat{\rho}_\mathrm{ss}) = 0$. 
Therefore, the Arnold tongue of synchronization vanishes, as shown in Fig.~\ref{fig:squareplot}(c). 

Surprisingly, the Arnold-tongue-like structure in the plot of the GP vanishes in the synchronization-blockade regime, too, even though the $\mathcal{S}(\hat{\rho}_\mathrm{ss})$ and $\gamma[\mathcal{P}]$ measure very different properties of the density matrix.
This result suggests a deeper connection between synchronization and the GP.

\subsection{Approximate analytic expression for the geometric phase}
\label{sec:approx_expr_gp}

To get a better understanding for the numerical results shown in Fig.~\ref{fig:squareplot}, we now derive approximate analytic expressions for the GP of a quantum limit-cycle oscillator whose quantization axis is slowly rotated as described by the QME~\eqref{eq:QME:LabFrame}.
In a first step, we use the fact that the rotation $\hat{R}(\alpha,t)$ is adiabatic, i.e., the timescale $2 \pi/\omega$ on which the direction of the quantization axis changes is much longer than any other timescale of the system. 
An approximate solution for $\hat{\rho}(t)$ can thus be obtained by calculating the steady state of the system for a fixed orientation of the quantization axis, and then rotating this steady state according to $\hat{R}(\alpha,t)$. 
This motivates us to introduce the frame which co-rotates with the quantization axis,
\begin{align}
	\hat{\chi}(t) = R^\dagger(\alpha,t) \hat{\rho}(t) \hat{R}(\alpha,t) \comma
	\label{eqn:Trafo:LabToAxisFrame}
\end{align}
which gives rise to the effective QME
\begin{align}
	\frac{\d}{\d t} \hat{\chi} = - i \commutator{\hat{H}_0 + \hat{H}_\mathrm{sig}(t) + \hat{H}_\mathrm{axis}}{\hat{\chi}} + \sum_{j=1}^M \mathcal{D} [ \hat{\Gamma}_j ]\hat{\chi} \comma  
	\label{eq:QME:AxisFrame}
\end{align}
where $\hat{H}_\mathrm{axis} = - i \hat{R}^\dagger(\alpha,t) \partial_t \hat{R}(\alpha,t) = - \omega \vec{n}(\alpha) \cdot \vec{\hat{S}}$ is the correction term to the Hamiltonian due to the slow rotation of the quantization axis. 
It causes a small tilt of the effective quantization axis in the co-rotating frame, $\omega_0 \hat{S}_z \to [\omega_0 - \omega \cos\alpha] \hat{S}_z - \omega \sin(\alpha) \hat{S}_x$.

Naively, one may now attempt to simplify Eq.~\eqref{eq:QME:AxisFrame} by switching to a rotating frame with respect to the signal, and by performing a rotating-wave approximation. 
However, this approach eliminates the $\hat{S}_x$ correction to the quantization axis and leads to incorrect results. 
To preserve this term, we first diagonalize the modified Hamiltonian $\hat{H}_0 + \hat{H}_\mathrm{axis}$ to leading order in $\omega/\omega_0$ using a Schrieffer-Wolff transformation 
\begin{align}
	\hat{\chi}_\mathrm{SW} &= e^{\hat{W}} \hat{\chi} e^{-\hat{W}} 
	\label{eqn:Trafo:AxisToSWFrame}
\end{align}
with the generator
\begin{align}
	\hat{W} &= - \frac{i \omega \sin \alpha}{\omega_0 - \omega \cos \alpha} \hat{S}_y \fullstop
\end{align}
The QME for the density matrix $\hat{\chi}_\mathrm{SW}$ in this new frame is
\begin{align}
	\frac{\d}{\d t} \hat{\chi}_\mathrm{SW} 
		&= - i \commutator{e^{\hat{W}} \left( \hat{H}_0 + \hat{H}_\mathrm{axis} + \hat{H}_\mathrm{sig}(t)\right) e^{- \hat{W}}}{\hat{\chi}_\mathrm{SW}} \nonumber\\
		&\phantom{=}\ + \sum_{j=1}^M \mathcal{D}\left[ e^{\hat{W}} \hat{\Gamma}_j e^{- \hat{W}} \right] \hat{\chi}_\mathrm{SW} \fullstop 
		\label{eq:QME:SWframe}
\end{align}
The Schrieffer-Wolff transformation rotates the spin basis states such that $\hat{H}_0 + \hat{H}_\mathrm{axis}$ becomes diagonal up to quadratic corrections,
\begin{align}
	e^{\hat{W}} \left( \hat{H}_0 + \hat{H}_\mathrm{axis} \right) e^{- \hat{W}}
	&= \left( \omega_0 - \omega \cos \alpha \right) \hat{S}_z + \mathcal{O}\left( \frac{\omega^2}{\omega_0^2} \right) \comma
\end{align}
whereas the signal acts now along a combination of the $x$ and $z$ directions,
\begin{align}
	e^{\hat{W}} \hat{H}_\mathrm{sig}(t) e^{- \hat{W}}
	&= T \left( \hat{S}_x - \frac{\omega \sin \alpha}{\omega_0 - \omega \cos\alpha} S_z \right) \cos(\omtilde t + \phitilde) \nonumber \\
	&\phantom{=}\ + \mathcal{O}\left( \frac{\omega^2}{\omega_0^2} \right) \fullstop
\end{align}
We can now finally switch to a frame rotating at the signal frequency $\omtilde$, 
\begin{align}
	\hat{\chi}_\mathrm{rot} &= \hat{U}^\dagger(t) \hat{\chi}_\mathrm{SW} \hat{U}(t) \comma 
	\label{eqn:Trafo:SWtoRWAframe}\\
	\hat{U}(t) &= \exp (-i \omtilde t \hat{S}_z) \comma 
\end{align}
and perform a rotating-wave approximation. 
The resulting QME is
\begin{align}
	\frac{\d}{\d t} \hat{\chi}_\mathrm{rot} &= - i \commutator{\hat{H}}{\hat{\chi}_\mathrm{rot}} + \sum_{j=1}^M \mathcal{D}\left[ \hat{\Gamma}_j \right] \hat{\chi}_\mathrm{rot} \comma 
	\label{eq:QME:RWAframe} \\
	\hat{H} &= \left(\omega_0 - \omtilde + \omega \cos\alpha \right) \hat{S}_z + \frac{T}{4} \left( e^{-i \phitilde} \hat{S}_+ + e^{i \phitilde} \hat{S}_- \right) \comma
\end{align}
which differs from the naive approach to simplify Eq.~\eqref{eq:QME:AxisFrame} outlined above by the fact that $\hat{\chi}_\mathrm{rot}$ is given in a basis which is rotated by $e^{\hat{W}}$ compared to the basis of $\hat{\chi}$ in Eq.~\eqref{eq:QME:AxisFrame}.
In this way, we retain information on the $\hat{S}_x$ term in $\hat{H}_\mathrm{axis}$. 

The steady state $\hat{\chi}_\mathrm{rot,ss}$ of Eq.~\eqref{eq:QME:RWAframe} has the same form as shown in Eqs.~\eqref{eq:sigmasteadystate} to~\eqref{eq:vdP:c0m}, except that the detuning now contains a correction term due to $\hat{H}_\mathrm{axis}$, $\Delta  \to \Delta - \omega \cos\alpha = \tilde{\omega} - \omega_0 - \omega \cos\alpha$.

\begin{figure}
    \centering
    \includegraphics[width=0.45\textwidth]{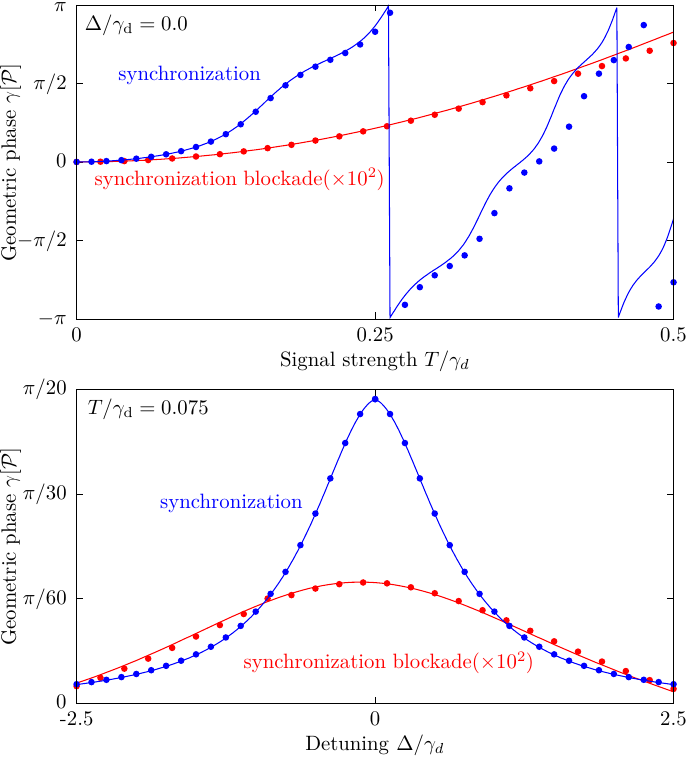}
    \caption{
    Linecuts through the different regions shown in Fig.~\ref{fig:linecuts} (data points), compared with the approximate analytic expressions for the GP from App.~\ref{sec:App:ApproximateExpressionGP} (solid lines). 
    The red dots and lines correspond to the synchronization-blockade parameters of Fig.~\ref{fig:linecuts}(d) and are magnified by a factor of $10^2$. 
    The blue dots and lines correspond to the parameters of Fig.~\ref{fig:linecuts}(b). 
    }
    \label{fig:linecuts}
\end{figure}

Assuming the populations $\{p_{+1}, p_0, p_{-1}\}$ to be nondegenerate, we can now diagonalize $\hat{\chi}_\mathrm{rot,ss}$ perturbatively in the signal strength $T$ and find the eigenvectors
\begin{align}
	\ket{p_{+1}} &\propto \left( 1 , \frac{T c_{+1,0}^*}{p_1 - p_0} , 0 \right)^\top + \mathcal{O}(T^2) \comma \\
	\ket{p_0} &\propto \left( -\frac{T c_{+1,0}}{p_1 - p_0}, 1, \frac{T c_{0,-1}^*}{p_0 - p_{-1}} \right)^\top + \mathcal{O}(T^2)\comma \\
	\ket{p_{-1}} &\propto \left( 0 , - \frac{T c_{0,-1}}{p_0 - p_{-1}} , 1 \right)^\top + \mathcal{O}(T^2) \fullstop
\end{align}
Their corresponding eigenvalues $p_{+1}$, $p_0$, and $p_{-1}$ remain unchanged up to corrections of $\mathcal{O}(T^2)$.
To obtain the lab-frame density matrix $\hat{\rho}$, we now undo the transformations~\eqref{eqn:Trafo:SWtoRWAframe}, \eqref{eqn:Trafo:AxisToSWFrame}, and~\eqref{eqn:Trafo:LabToAxisFrame}. 
The frequency $\omega$ appears twice in these transformations, namely, as the small expansion parameter $\omega/\omega_0$ in the Schrieffer-Wolff transformation, and as the potentially large rotation angle $\omega t$ of the quantization axis. 
To separate these different roles of $\omega$ clearly, we introduce two new parameters for the backtransformation, $\omega \to \omega_\mathrm{SW}$ in Eq.~\eqref{eqn:Trafo:AxisToSWFrame} and $\omega \to \omega_\mathrm{R}$ in Eq.~\eqref{eqn:Trafo:LabToAxisFrame}, such that we can work perturbatively in $\omega_\mathrm{SW}/\omega_0$ while keeping all orders of $\omega_\mathrm{R}/\omega_0$,
\begin{align}
	\ket{\phi_m(t)} &= \hat{R}(t,\alpha,\omega_\mathrm{R}) [\hat{\mathds{1}} - \hat{W}(\omega_\mathrm{SW})] \hat{U}(t) \ket{p_m} \comma
\end{align}
where $m \in \{+1,0,-1\}$.
At the end of the calculation, we will set $\omega_\mathrm{R} = \omega_\mathrm{SW} = \omega$.
The full expressions for the eigenvectors $\ket{\phi_m(t)}$ are lengthy and intransparent.
Simplified expressions for the overlaps $\bra{\phi_m(0)}\ket{\phi_m(\tau)}$ and the integrals $-\int_0^\tau \d t \bra{\phi_m(t)}\ket{\dot{\phi}_m(t)}$ entering Eq.~\eqref{eq:GPtong} are given in App.~\ref{sec:App:ApproximateExpressionGP}. 
For cyclic evolution, $\tau \to 2 \pi/\omega$, they reduce to $\gamma[\mathcal{P}] = \arg(z)$ with
\begin{align}
	z &= p_{+1} \exp \left[+2 \pi i \cos(\alpha) + \sqrt{2} i T \sin(\alpha) \frac{\omega}{\omega_0} \frac{\Im c_{+1,0}}{p_1 - p_0} \right] \nonumber \\
		&+ p_0 \exp \left[ \sqrt{2} i T \sin (\alpha) \frac{\omega}{\omega_0} \left( \frac{\Im c_{0,-1}}{p_0 - p_{-1}} - \frac{\Im c_{+1,0}}{p_1 - p_0} \right) \right] \nonumber\\
		&+ p_{-1} \exp \left[ -2 \pi i \cos(\alpha) - \sqrt{2} i T \sin (\alpha) \frac{\omega}{\omega_0} \frac{\Im c_{0,-1}}{p_0 - p_{-1}} \right] \fullstop 
	\label{eq:GP:approximateCyclic}
\end{align}
If no signal is applied, $T \to 0$, this result reduces to our guess for the GP of an unperturbed quantum limit-cycle oscillator given in Eq.~\eqref{eq:GP:CyclicNoSignal}.
These analytical formulas are in excellent agreement with the numerical results shown in Fig.~\ref{fig:squareplot}(b) and (d) if $\omega$ and $T$ are small, as demonstrated in Fig.~\ref{fig:linecuts}.

We stress that the derivation presented in this section and the result~\eqref{eq:GP:CyclicNoSignal} are not specific to the quantum van der Pol oscillator considered in our numerical examples. 
Any spin-$1$ limit-cycle oscillator subject to a semiclassical signal of the form~\eqref{eq:Hsig} will have a density matrix of the form~\eqref{eq:sigmasteadystate}~\cite{Koppenhoefer2019}.
Only the specific formulas for the populations $p_m$ and the coherences $T c_{m,m'}$ will differ depending on the dissipators $\hat{\Gamma}_k$ stabilizing the limit cycle.
The results can thus be directly applied to other limit-cycle oscillators (as long as all nonzero populations $p_k$ are distinct), and they can be easily generalized to other spin numbers $S$.

\subsection{Interpretation of the Arnold tongue of the geometric phase}
The derivation of the approximate expression~\eqref{eq:GP:approximateCyclic} for the GP of a quantum limit-cycle oscillator provides an intuitive understanding of the results shown in Fig.~\ref{fig:squareplot}. 
As shown in Fig.~\ref{fig:spin1_rot}, the quantization axis slowly rotates on the surface of a cone. 
In the absence of an external signal, $T = 0$, each eigenstate $\ket{S,m}$ traces out a path with a solid angle $2 \pi (1 - \cos\alpha) |m|$ subtended from the origin of the Bloch sphere, as shown by the red curve in Fig.~\ref{fig:precession_plot}.
The time-dependent signal Hamiltonian $\hat{H}_\mathrm{sig}(t)$ in Eq.~\eqref{eq:QME:LabFrame} tries to tilt the states $\ket{S,m}$ away from the instantaneous quantization axis and causes them to precess, as shown by the blue curve in Fig.~\ref{fig:precession_plot}. 
The action of the drive is counteracted by the dissipative limit-cycle stabilization mechanism, which attempts to relax the system to a state without any coherences in the basis defined by the instantaneous direction of the quantization axis.
In a reference frame that corotates with $\hat{H}_\mathrm{sig}(t)$ about the instantaneous quantization axis [see Eq.~\eqref{eqn:Trafo:SWtoRWAframe}], the density matrix thus acquires constant coherences, as shown in Eq.~\eqref{eq:sigmasteadystate}. 

Since their magnitude increases with the signal strength $T$ and decreases with increasing detuning $\abs{\Delta}$,
the GP changes in an Arnold-tongue-like region around resonance.

In the case of cyclic evolution shown in Eq.~\eqref{eq:GP:approximateCyclic}, only the
imaginary part of the coherences modifies the phase factor of each eigenstate $\ket{\phi_m}$.
For noncyclic evolution, the coherences also change the overlap $\bra{\phi_m(0)}\ket{\phi_m(t)}$ between eigenstates, see App.~\ref{sec:App:ApproximateExpressionGP} for details. 
In both cases, however, the functional dependence of $\gamma[\mathcal{P}]$ on the coherences $T c_{m,m'}$ is more complex than the simple sum of coherences encountered in the synchronization measure $\mathcal{S}(\hat{\rho})$.
Therefore, the suppression of the Arnold tongue of the GP for parameters in the synchronization-blockade regime must have a different origin than the destructive interference of coherences that causes the Arnold tongue of synchronization to vanish.

The surprising disappearance of the Arnold tongue of the GP in Fig.~\ref{fig:squareplot}(d) can be traced back to a more general suppression of coherences in the synchronization blockade regime.
As shown in Eqs.~\eqref{eq:vdP:cp0} and~\eqref{eq:vdP:c0m}, 
the coherences are in general different functions of the gain and dissipation rates $\gamma_{\mathrm{g}}$ and $\gamma_{\mathrm{d}}$.
Focusing on the resonant case $\Delta = 0$, we find
\begin{align}
	c_{+1,0} 
		&\propto \frac{4 \gamma_\mathrm{g} \gamma_\mathrm{d} + 3 \sqrt{2} \gamma_\mathrm{g}(\gamma_\mathrm{d} - \gamma_\mathrm{g})}{12 \gamma_\mathrm{g} (3 \gamma_\mathrm{d} + \gamma_\mathrm{g})(\gamma_\mathrm{d} + \gamma_\mathrm{g})} \nonumber \\
		&\approx - \frac{1}{2 \sqrt{2} \gamma_\mathrm{d}} \frac{1}{\gamma} + \mathcal{O}\left( \gamma^{-2} \right)  \comma 
		\label{eqn:cp10approx} \displaybreak[1]\\
	c_{0,-1} 
		&\propto \frac{\gamma_\mathrm{d}}{3 \sqrt{2} \gamma_\mathrm{g} (3 \gamma_\mathrm{d} + \gamma_\mathrm{g})} \nonumber \\
		&\approx + \frac{1}{3 \sqrt{2} \gamma_\mathrm{d}} \frac{1}{\gamma^2} + \mathcal{O}\left( \gamma^{-3} \right) \comma
		\label{eqn:c0m1approx}
\end{align}
where we introduced the ratio $\gamma = \gamma_\mathrm{g}/\gamma_\mathrm{d}$ and considered the limit of large $\gamma$. 
For generic values of $\gamma_\mathrm{g}$ and $\gamma_\mathrm{d}$, the coherences $T c_{+1,0}$ and $T c_{0,-1}$ will have different magnitudes and will not interfere destructively.
However, the expressions in the first lines of Eqs.~\eqref{eqn:cp10approx} and~\eqref{eqn:c0m1approx} show that the coherences tend to zero with increasing dissipation rates $\gamma_\mathrm{d}$ or $\gamma_\mathrm{g}$ because the limit-cycle stabilization scheme dominates over the influence of the signal. 
Therefore, one can find specific combinations of dissipation rates for which the destructive interference occurs, e.g., by increasing the ratio $\gamma_\mathrm{g}/\gamma_\mathrm{d}$.
This fine-tuning of the ratio $\gamma_\mathrm{g}/\gamma_\mathrm{d}$, however, comes at the cost of an overall reduction of the magnitude of the coherences (about an order of magnitude for our parameters), which leads to a slower change of the GP when $T$ or $\Delta$ are changed. 
This effect causes the disappearance of the Arnold tongue of the GP in Fig.~\ref{fig:squareplot}(d).

\section{Conclusion}
\label{sec:conclusion}
Based on the kinematic approach to the geometric phase (GP) proposed by Tong \emph{et al.}~\cite{Tong2004}, we have developed a numerically stable algorithm to calculate the GP in an open quantum system. We used it to demonstrate the existence of a GP in a spin-$1$ implementation of a quantum vdP oscillator whose quantization axis is slowly rotated. 
We have shown that if the quantum vdP oscillator is synchronized to an external signal, the GP plotted as a function of the detuning and signal strength exhibits a structure similar to the Arnold tongue of synchronization: 
the GP changes strongly in a roughly triangular region around resonance whose width increases with increasing signal strength.
Surprisingly, this Arnold tongue of the GP vanishes if the system is in a parameter regime where an interference-based quantum synchronization blockade occurs.

These striking similarities between the Arnold tongue of the synchronization measure $\mathcal{S}(\hat{\rho})$ and the structure of the GP naturally lead to the question if there is a deeper connection between GPs and quantum synchronization. 
For instance, could quantum synchronization be an indicator of a nonzero GP or vice versa? 
In general, a nonzero GP does \emph{not} imply that a quantum system is synchronized because, even in the absence of an external signal, the quantum vdP limit-cycle oscillator shows a GP (see Sec.~\ref{sec:quantumvdP:Undriven}) but it is clearly not synchronized. 
Moreover, GPs occur even in completely unitary evolution \cite{Berry1984}, i.e., in quantum systems that are no limit-cycle oscillators at all. 

Conversely, the numerical data presented in Sec.~\ref{sec:quantumvdP:driven} suggests that a nonzero synchronization measure $\mathcal{S}(\hat{\rho})$ could be an indicator of changes in the GP relative to its value in an unperturbed limit-cycle oscillator.

Using perturbation theory in the small frequency $\omega$ of the rotation of the quantization axis and in the small signal strength $T$, we have derived an approximate analytical expression for the GP, which reveals that the mechanism leading to the suppression of the GP (namely, a suppression of the coherences compared to a regime of regular synchronization) is different from the mechanism leading to a suppression of the synchronization measure (namely, destructive interference of the coherences).

Despite these differences, the synchronization measure $\mathcal{S}(\hat{\rho})$ and the GP show qualitatively the same behavior in a quantum vdP oscillator. 
It is an exciting direction for further research to understand if this is a generic feature that holds for arbitrary limit-cycle oscillators and external signals. 
Since the assumptions in our derivation of a perturbative analytic formula for the GP in Sec.~\ref{sec:approx_expr_gp} are very general, the same technique could be applied to other quantum limit-cycle oscillators to address this open question.

Kepler \emph{et al.}~\cite{Kepler1991a,Kepler1991b} developed a general approach to the GP in classical limit-cycle systems, but the deformations of the classical limit cycle they analyzed differ from the rotation of the quantization axis we considered here.
It would therefore be interesting to connect and compare these results by analyzing the classical equivalent of a quantum vdP oscillator whose quantization axis slowly rotates.

\begin{acknowledgments}
We would like to thank P.\ Sekatski and E.\ Sj\"oqvist for discussions, and 
acknowledge important contributions by L.\ Fricker and A.\ Roulet to early stages of this project.
This work was financially supported by the Swiss National Science
Foundation (SNSF) (grants No. 185902 (NCCR QSIT: Quantum Science and Technology) and No. 200481). A.D. acknowledges financial support by the QCQT PhD school. 
\end{acknowledgments}

\appendix

\section{Interferometric measurement of the geometric phase}
\label{sec:App:mzi_measurement}

In this appendix, we comment on the possibility to measure the  GP of a mixed state undergoing nonunitary evolution in an interferometric measurement. 

\begin{figure}
	\centering
	\def\svgwidth{\columnwidth}
    \includegraphics[width=0.45\textwidth]{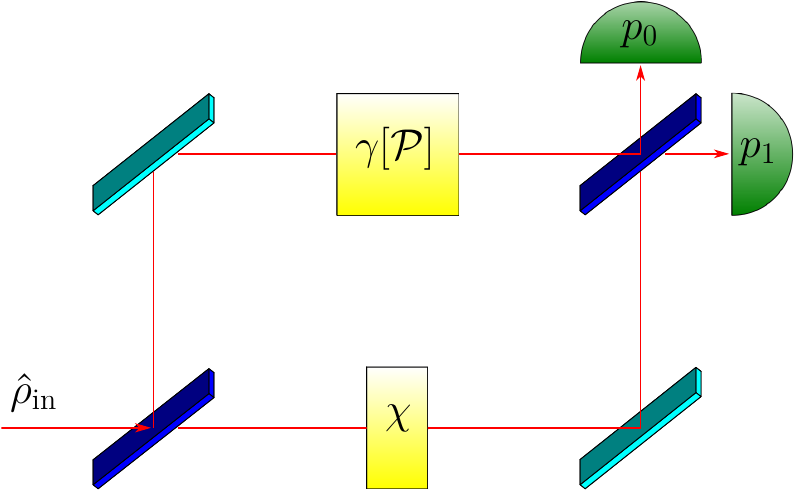}
	\caption{
		Mach-Zehnder interferometer (MZI) setup to measure the GP of a mixed state undergoing unitary evolution.
		The state in the upper arm undergoes adiabatic unitary evolution along the path $\mathcal{P}$ and acquires a  
        GP $\gamma[\mathcal{P}]$ 
        whereas the lower arm acquires only a $U(1)$ reference phase $\chi$. 
		The two detectors (green) measure the probabilities of the state exiting in the two output ports of the second beam splitter.
	}
	\label{fig:MZI}
\end{figure}

In the case of a \emph{pure} initial state and \emph{unitary} evolution, the GP is uniquely defined and can be measured in an interferometric measurement, as shown in Fig.~\ref{fig:MZI}.
The first beam splitter transforms the initial pure state into a superposition of pure states propagating along the two arms of the Mach-Zehnder interferometer (MZI).
The state in the upper arm undergoes adiabatic unitary evolution along the desired path $\mathcal{P}$ and acquires a GP $\gamma[\mathcal{P}]$, whereas the state in the lower arm acquires only a controllable $U(1)$ reference phase $\chi$. 
Having passed through the two arms, the states interfere at the second beam splitter and the probabilities $p_0$ and $p_1$ of finding the system in the two output ports are measured. These probabilities depend on the relative phase $\gamma[\mathcal{P}] - \chi$ between the two arms and show an interference profile of the form 
\begin{align}
	p_{0,1} = \frac{1}{2} \left[ 1 \pm \nu \cos (\chi - \gamma[\mathcal{P}]) \right] \fullstop
	\label{eqn:FiringProbabilities}
\end{align}
The dynamical phase can be eliminated from the measurement by enforcing the parallel-transport condition in the upper arm or by choosing a specific path along which the dynamical phase vanishes. 
Other possibilities include cancelling the dynamical phase using spin-echo techniques, 
or comparing different paths where the relative signs between the dynamical and GPs differ. 
One can then use the controllable phase $\chi$ to map out Eq.~\eqref{eqn:FiringProbabilities} and determine the GP $\gamma[\mathcal{P}]$.

The GP of a mixed state is defined via a purification of the state in a larger Hilbert space~\cite{Uhlmann1986,Sjoqvist2000,Singh2003}.
Since a given mixed state can be purified in many different ways, one may worry that the GP can no longer be uniquely defined.
However, Sj\"oqvist \emph{et al.}~\cite{Sjoqvist2000} showed that, for \emph{mixed} states undergoing \emph{unitary} evolution, the GP can still be measured in the MZI setup shown in Fig.~\ref{fig:MZI} and turns out to be the statistical average of the GPs of the pure eigenstates of the density matrix. 
Such an interferometric measurement has been experimentally demonstrated in an NMR system~\cite{Du2003}.

In the following, we calculate the measurement probabilities of a mixed state in a MZI undergoing \emph{nonunitary} evolution and show that the interferometric measurement of the GP of mixed state \textit{cannot} be extended to nonunitary evolution. 

Analogously to the treatment of the unitary case in~\cite{Sjoqvist2000}, we model the MZI in a combined Hilbert space $\mathcal{H}_B \otimes \mathcal{H}_S$, where $\mathcal{H}_B = \{ \ket{0}^B,\ket{1}^B \}$ encodes the two different paths the system can take inside the interferometer, and $\mathcal{H}_S$
is the Hilbert space of the system.
The beam splitters implement the unitary transformation 
\begin{align}
	\hat{U}_\mathrm{BS} = \frac{1}{\sqrt{2}} \left[ \ket{0}\bra{0}^B + \ket{0}\bra{1}^B + \ket{1}\bra{0}^B - \ket{1}\bra{1}^B \right] \comma
\end{align}
where the superscript $B$ denotes that this operator acts only on the subspace $\mathcal{H}_B$. 
After the first beam splitter, a phase shift $\chi$ is applied in the $\ket{0}^B$ path,
\begin{align}
	\hat{U}_\mathrm{PS}^B = e^{i \chi} \ket{0}\bra{0}^B + \ket{1}\bra{1}^B \fullstop
	\label{eq:UPS}
\end{align}
Along the $\ket{1}^B$ path, the system evolves in its Hilbert space $\mathcal{H}_S$ under a QME of the same form as Eq.~\eqref{eq:QME}, 
\begin{align}
\dv{t} \hat{\rho}^S(t) 
	&= -i \qty[\hat{H}^S(t),\hat{\rho}^S(t)] + \sum_{j=1}^M \mathcal{D}\qty[\hat{\Gamma}^S_j(t)] \hat{\rho}^S(t) \nonumber \\
	&= \mathcal{L}^S(t) \hat{\rho}^S(t) \comma
	\label{eq:QMESystemSubspace}
\end{align}
with a Hamiltonian $\hat{H}^S(t)$ and a set of Lindblad operators $\{ \hat{\Gamma}^S_j(t)\}$. 

We now lift these operations from the individual Hilbert spaces $\mathcal{H}_S$ and $\mathcal{H}_B$ into the combined Hilbert space $\mathcal{H}_B \otimes \mathcal{H}_S$, 
\begin{align}
	\hat{U}_\mathrm{BS} &= \hat{U}_\mathrm{BS}^B \otimes \hat{\1}^S \comma \\
	\hat{H}(t) &= \ket{1} \bra{1}^B \otimes \hat{H}^S(t) - \frac{\chi}{\tau} \ket{0} \bra{0}^B \otimes \hat{\1}^S \comma 
	\label{eq:combined_system_H_MZI}\\
	\hat{\Gamma}_j(t) &= \ket{1} \bra{1}^B \otimes \hat{\Gamma}_j^S(t) \comma
\end{align}
where $\hat{\1}^S$ ($\hat{\1}^B$) denotes the identity operator on $\mathcal{H}_S$ ($\mathcal{H}_B$). 
The second term in the Hamiltonian~\eqref{eq:combined_system_H_MZI} is the generator of the phase-shift transformation~\eqref{eq:UPS}.
It ensures that the system has acquired a phase shift $\chi$ in the $\ket{0}^B$ arm by the time $\tau$ when it exits the interferometer. 
Inside the interferometer, the time evolution of the entire system is governed by the following QME for the density matrix $\hat{\rho}(t)$ defined on $\mathcal{H}_\mathrm{B} \otimes \mathcal{H}_S$, 
\begin{align}
	\dv{t} \hat{\rho}(t) = -i \qty[\hat{H}(t),\hat{\rho}(t)] + \sum_i\mathcal{D}\qty[\hat{\Gamma}_i(t)] \hat{\rho}(t) \fullstop
\end{align}

To calculate an expression for the phase shift and the visibility in Eq.~\eqref{eqn:FiringProbabilities}, we consider the following product state entering the MZI,
\begin{align}
	\hat{\rho}_{0} 
	= \ket{0} \bra{0}^B \otimes \hat{\rho}^S(0) \comma
\end{align}
which is transformed into $\hat{\rho}_1 = \hat{U}_\mathrm{BS} \hat{\rho}_0 \hat{U}_\mathrm{BS}^\dagger$ at the first beam splitter. 
The time evolution after the first beam splitter is given by
\begin{align}
	\dv{t} \hat{\rho}(t) = 
	\frac{1}{2} \Big[ \begin{aligned}[t]
		&- i \ket{1}\bra{0}^B \otimes \hat{H}_\mathrm{eff}^S(t) \hat{\rho}^S(t) \\ 
		&+ i \ket{0}\bra{1}^B \otimes \hat{\rho}^S(t) \hat{H}_\mathrm{eff}^{S\dagger}(t) \\
		&+ \ket{1}\bra{1}^B \otimes \mathcal{L}^S(t) \hat{\rho}^S(t) \Big] \comma 
		\end{aligned}
	\label{eq:QME_general}
\end{align}
where we defined the effective non-Hermitian Hamiltonian
\begin{align}
	\hat{H}_\mathrm{eff}^S(t) &= \hat{H}^S(t) - \frac{i}{2} \sum_{j=1}^M \hat{\Gamma}_j^{S\dagger} \hat{\Gamma}_j^S + \frac{\chi}{\tau} \fullstop
\end{align}
The time evolution acts separately on the subspaces spanned by the populations and coherences of $\mathcal{H}_B$, such that we can formally solve it by introducing the effective time-evolution operator
\begin{align}
	\hat{U}_\mathrm{eff}^S(t) 
		&= \mathcal{T} \exp \left[ - i \int_0^t \d t' \hat{H}_\mathrm{eff}^S(t') \right] \nonumber \\
		&= e^{-i \chi t/\tau} \tilde{U}_\mathrm{eff}^S(t) \comma
\end{align}
where $\mathcal{T}$ denotes time ordering, as well as the effective time-evolution superoperator $\mathcal{U}^S_\mathrm{eff}(t)$ solving Eq.~\eqref{eq:QMESystemSubspace}.
The explicit form of $\mathcal{U}^S_\mathrm{eff}(t)$ is irrelevant in the following, and we only need its property to preserve the trace of $\hat{\rho}^S(t)$. 
With these definitions, a formal solution for the state between the two beam splitters is
\begin{align}
	\hat{\rho}_2(t) 
		=
  \begin{aligned}[t]
        \frac{1}{2} &\left(\ket{1}\bra{0}^B \otimes \hat{U}^S_\mathrm{eff}(t) \hat{\rho}^S(0) \right.\\
		&+ \ket{0}\bra{1}^B \otimes \hat{\rho}^S(0) \hat{U}^{S\dagger}_\mathrm{eff}(t) \\
		&+ \left.\ket{1}\bra{1}^B \otimes \mathcal{U}^S_\mathrm{eff}(t) \hat{\rho}^S(0) 
		+ \ket{0}\bra{0}^B \otimes \hat{\rho}^S(0)\right) \fullstop
  \end{aligned} 
\end{align}
The final state exiting the interferometer after the second beam splitter is $\hat{\rho}_3 = \hat{U}_\mathrm{BS} \hat{\rho}_2(\tau) \hat{U}_\mathrm{BS}^{\dagger}$ and the detection probabilities of the two detectors are 
\begin{align}
	p_0 &= \bra{0}_B \Tr_S \qty[\hat{\rho}_3]\ket{0}_B \nonumber \\ 
		&= \frac{1}{4} \Tr_S \Big[ \begin{aligned}[t]
			&\mathcal{U}^S_\mathrm{eff}(\tau) \hat{\rho}^S(0) + \hat{\rho}^S(0) \\
			+ &\hat{U}_\mathrm{eff}^S(\tau) \hat{\rho}^S(0) + \hat{\rho}^S(0) \hat{U}_\mathrm{eff}^{S\dagger} \Big] 
			\end{aligned} \nonumber \\
		&= \frac{1}{2} \left[ 1 + \nu(\tau) \cos\qty[\chi - \gamma(\tau)] \right] \comma \\
	p_1 &= 1 - p_0 \comma
\end{align}
where we used $\Tr_S \left[ \hat{U}^{S}_\mathrm{eff}(\tau) \hat{\rho}^S(0) \right] = \Tr_S \left[ \hat{\rho}^S(0) \hat{U}^{S\dagger}_\mathrm{eff}(\tau) \right]^*$ and defined the visibility $\nu(\tau)$ and phase shift $\gamma(\tau)$ as follows.
\begin{align}
	\nu(\tau) &= \abs{\Tr_S \left[ \tilde{U}_\mathrm{eff}^S(\tau) \hat{\rho}^S(0) \right]} \comma 
	\label{eq:MZI:visibility}\\
	\gamma(\tau) &= \arg \left( \Tr_S \left[ \tilde{U}_\mathrm{eff}^S(\tau) \hat{\rho}^S(0) \right] \right) \fullstop
	\label{eq:MZI:phase}
\end{align}

The presence of the non-Hermitian term $-(i/2) \sum_{j=1}^M \hat{\Gamma}_j^{S\dagger} \hat{\Gamma}_j^S$ in $\hat{H}_\mathrm{eff}^S(t)$ implies that $\nu(\tau)$ decays to zero with increasing evolution time $\tau$. 
For instance, for the spin-$1$ quantum vdP oscillator introduced in Sec.~\ref{sec:gp_quantum_lc} and ignoring the time-dependent rotation of the quantization axis for a moment, the effective unitary operator is 
\begin{align}
	\hat{U}_{\mathrm{eff}}(\tau) = e^{-i\chi -i \hat{H}_0 \tau-\frac{1}{2}(\hat{\Gamma}_1^{\dagger} \hat{\Gamma}_1+\hat{\Gamma}_2^{\dagger}\hat{\Gamma}_2)\tau}\comma
\end{align}
where the Hamiltonian $\hat{H}_0$ and the Lindblad operators $\hat{\Gamma}_1$, $\hat{\Gamma}_2$ are defined in Eqs.~\eqref{eq:H0} to~\eqref{eq:jump_operators:2}. 
Hence, the visibility decays proportional to $ e^{-\text{min}(\gamma_{\mathrm{g}},\gamma_{\mathrm{d}}) \, \tau}$. 
For large damping and gain rates or for long times $\tau$, the visibility of the interference pattern tends to zero, $\nu \to 0$, and detection of any phase shift becomes impossible.
Moreover, $\gamma(\tau)$ has a very different form than the definition of a GP in nonunitary evolution, Eq.~\eqref{eq:GPtong}, such that $\gamma(\tau) \neq \gamma[\mathcal{P}]$.

In conclusion, for mixed states and nonunitary time evolution, the interferometric measurement considered here cannot be used to determine the GP.
Since the GP for nonunitary evolution is defined via a purification procedure \cite{Uhlmann1986,Tong2004}  and since the choice of the purification matters \cite{Rezakhani2006,Zhu2011}, one has to enforce this purification during the interferometric measurement. 
This implies that one has to perform an interferometric measurement using \emph{unitary} evolution of the combined system and ancilla~\cite{Aberg2007,Cucchietti2010,Zhu2011,Sjoqvist2020}, which is experimentally very demanding for large quantum systems.

The phase $\gamma(\tau)$ defined in Eq.~\eqref{eq:MZI:phase} is related to a definition of geometric phases in open quantum systems using a quantum-jump unraveling of the dissipative dynamics \cite{Carollo2003}. 
A geometric phase can then be defined by adding up the phase changes at quantum jumps and during nonunitary time evolution between quantum jumps, but the value of the phase still depends on the chosen unraveling \cite{Bassi2006}.
In our calculation, the effective time-evolution operator $\tilde{U}_\mathrm{eff}^S(\tau)$ describes the dynamics of the system between two quantum-jump events and $\gamma(\tau)$ thus corresponds to the special (and rare) trajectory where no jumps occur during the entire duration $\tau$.

\begin{widetext}
\section{Geometric phase of a dephasing qubit}
\label{sec:App:QubitDephasingBenchmark}

\begin{figure}
    \centering
    \includegraphics[width=0.45\textwidth]{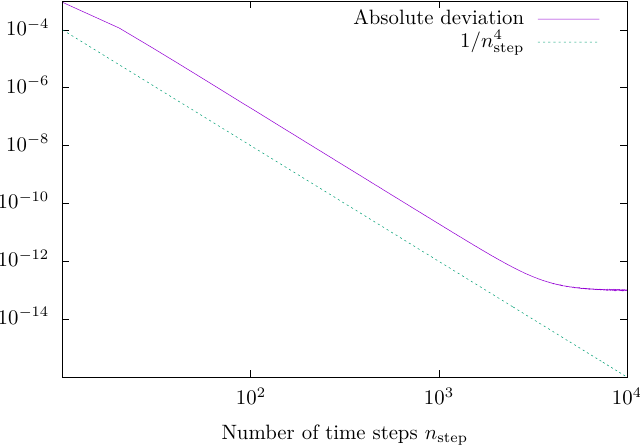}
    \caption{
    Deviation of the numerically calculated GP from the exact solution for a qubit undergoing pure dephasing, modeled by the quantum master equation~\eqref{eq:App:QubitDephasingQME}.
    The GP is computed using Alg.~\ref{alg:gp_algorithm} for a fixed evolution time $\tau =  2\pi/\eta$ and different numbers $\nstep$ of time steps. The dashed line indicates a $1/\nstep^4$ scaling. 
    The parameters are $\Lambda/\eta = 0.2$ and $ \theta_0 = \pi/4$.}    
    \label{fig:qubit_conv}
\end{figure}

In this appendix, we demonstrate the convergence of the numerical algorithm introduced in Sec.~\ref{sec:NumericalAlgorithm} of the main text on a simple analytically solvable example.
Tong \emph{et al.}~\cite{Tong2004} considered the GP of a two-level system subject to pure dephasing, defined by the QME
\begin{align}
	\frac{\d}{\d t} \hat{\rho} = - i \left[ \frac{\eta}{2} \hat{\sigma}_z , \hat{\rho} \right] + \frac{\Lambda}{2} \mathcal{D}[\hat{\sigma}_z] \hat{\rho}~,
	\label{eq:App:QubitDephasingQME}
\end{align}
where $\hat{\sigma}_z$ denotes the Pauli $z$-matrix.
Starting in the initial state 
\begin{align}
	\hat{\rho}(0) = \frac{1 + \vec{r} \cdot \hat{\boldsymbol{\sigma}}}{2} \comma
\end{align}
where $\hat{\boldsymbol{\sigma}} = (\hat{\sigma}_x, \hat{\sigma}_z, \hat{\sigma}_z)^\top$ denotes the vector of Pauli matrices and $\vec{r} = (\sin \theta_0, 0, \cos \theta_0)^\top$, the GP at time $\tau > 0$ is given by

\begin{align}
	\gamma(\tau) 
		&= \arg \left[ e^{-i \eta \tau/2} \cos \left(\frac{\theta_\tau}{2}\right) \cos \left(\frac{\theta_0}{2}\right) + e^{+i \eta \tau/2} \sin \left(\frac{\theta_\tau}{2} \right) \sin \left(\frac{\theta_0}{2}\right) \right] \nonumber \\
		&\phantom{=}\ + \frac{\eta}{4 \Lambda} \ln \frac{(1-\cos\theta_0) \left( \sqrt{\cos^2(\theta_0) + \sin^2(\theta_0) e^{-2 \Lambda \tau}} + \cos \theta_0 \right)}{(1+\cos\theta_0) \left( \sqrt{\cos^2(\theta_0) + \sin^2(\theta_0) e^{-2 \Lambda \tau}} - \cos \theta_0 \right)} \comma
	\label{eq:qubit_GP_cyclic}
\end{align}
where $\theta_\tau = \left[ \arctan \left( e^{- \Gamma \tau}\tan \theta_0 \right) + \pi \right] \operatorname{mod} \pi$. 
For a total evolution time $\tau = 2 \pi / \eta$ and $\cos \theta_0 \geq 0$, this result simplifies to 
Eq.~(21) of~\cite{Tong2004}.
With this analytical formula for the GP at hand, we benchmarked the accuracy of our numerical algorithm described in Sec.~\ref{sec:NumericalAlgorithm}. 
We solved Eq.~\eqref{eq:App:QubitDephasingQME}  numerically, calculated the GP using Alg.~\ref{alg:gp_algorithm}, and compared the result with Eq.~\eqref{eq:qubit_GP_cyclic}. 
For all tested parameters $\eta$ and $\Lambda$, and for all initial conditions $\vec{r}$, we observed quartic convergence in the number of time steps, similar to the result shown in Fig.~\ref{fig:qubit_conv}.

\section{Approximate expression for the geometric phase}
\label{sec:App:ApproximateExpressionGP}

In this appendix, we provide simplified expressions for the overlaps $\bra{\phi_m(0)}\ket{\phi_m(\tau)}$ and the integrals $-\int_0^\tau \d t \bra{\phi_m(t)}\ket{\dot{\phi}_m(t)}$ entering Eq.~\eqref{eq:GPtong}. 
These results generalize Eq.~\eqref{eq:GP:approximateCyclic} of the main text.
Ignoring fast oscillating terms $\propto e^{\pm i \tilde{\omega} t}$, 
assuming a resonant drive, $\Delta = 0$, and taking the limit $\omega_\mathrm{SW} \to 0$, we find the following expressions for the overlap of the eigenstates of the density matrix $\hat{\rho}(t)$, 
\begin{align}
	\bra{\phi_{+1}(0)}\ket{\phi_{+1}(\tau)} 
		&= \left( \cos \frac{\omega \tau}{2} - i \cos \alpha \sin \frac{\omega \tau}{2} \right) \left[ \left( \cos \frac{\omega \tau}{2} - i \cos \alpha \sin \frac{\omega \tau}{2} \right) - \sqrt{2} i T \frac{c_{+1,0}}{p_{+1} - p_0} \sin \alpha \sin \frac{\omega \tau}{2}  \right] \comma \\
	\bra{\phi_0(0)}\ket{\phi_0(\tau)} 
		&= \Bigg[ \begin{aligned}[t]
		\cos^2 \alpha + \sin^2 \alpha \cos \omega \tau - \sqrt{2} i T \sin \alpha \sin \frac{\omega \tau}{2} \Big[ &- \left( \cos \frac{\omega \tau}{2} - i \cos \alpha \sin \frac{\omega \tau}{2} \right) \frac{c_{+1,0}^*}{p_{+1} - p_0} \\
		&+ \left( \cos \frac{\omega \tau}{2} + i \cos \alpha \sin \frac{\omega \tau}{2} \right) \frac{c_{0,-1}}{p_0 - p_{-1}} \Big] \Bigg] \comma
		\end{aligned} \\
	\bra{\phi_{-1}(0)}\ket{\phi_{-1}(\tau)} 
		&= \left( \cos \frac{\omega \tau}{2} + i \cos \alpha \sin \frac{\omega \tau}{2} \right) \left[ \left( \cos \frac{\omega \tau}{2} + i \cos \alpha \sin \frac{\omega \tau}{2} \right) + \sqrt{2} i T \frac{c_{0,-1}^*}{p_0 - p_{-1}} \sin \alpha \sin \frac{\omega \tau}{2}  \right] \fullstop 
\end{align}
Moreover, the integrals determining the phase factors in Eq.~\eqref{eq:GPtong} are given by
\begin{align}
	- \int_0^\tau \d t\, \bra{\phi_{+1}(t)}\ket{\dot{\phi}_{+1}(t)}
		&= +i \omega \tau \cos \alpha + \frac{\sqrt{2} i \omega T \sin \alpha}{\omega_0} \frac{\Im c_{+1,0}}{p_1 - p_0} \comma \\
	- \int_0^\tau \d t\, \bra{\phi_0(t)}\ket{\dot{\phi}_0(t)}
		&= \frac{\sqrt{2} i \omega T \sin \alpha}{\omega_0} \left( \frac{\Im c_{0,-1}}{p_0 - p_{-1}} - \frac{\Im c_{+1,0}}{p_1 - p_0} \right) \comma \\
	- \int_0^\tau \d t\, \bra{\phi_{-1}(t)}\ket{\dot{\phi}_{-1}(t)}
		&= -i \omega \tau \cos \alpha - \frac{\sqrt{2} i \omega T \sin \alpha}{\omega_0} \frac{\Im c_{0,-1}}{p_0 - p_{-1}}\fullstop
\end{align}
\end{widetext}

\bibliography{library.bib}

\end{document}